\documentclass[5p]{elsarticle}
\usepackage{graphicx}
\usepackage{epsfig,latexsym}
 
\newcommand{\be}{\begin{eqnarray}}
\newcommand{\ee}{\end{eqnarray}}

\makeindex

\begin{document}

\title{Detecting a First-Order Transition in the QCD Phase Diagram \\with Baryon-Baryon Correlations}

\author[rvt]{ \'Agnes M\'ocsy}
\ead{amocsy@pratt.edu}

\author[focal]{ Paul Sorensen}
\ead{prsorensen@bnl.gov}

\address[rvt]{Department of Mathematics and Science, Pratt Institute, Brooklyn, NY 11205, USA}
\address[focal]{Brookhaven National Laboratory, Upton, NY 11973 USA}

\begin{abstract}
 We suggest baryon-baryon correlations as an experimentally
 accessible signature for a first-order phase transition between a
 baryon-rich phase, like quarkyonic, and a baryon-suppressed hadronic
 phase in the QCD phase diagram. We examine the consequences of
 baryon-rich bubble formation in an expanding medium and show how the
 two-particle correlations vary in the transverse and longitudinal
 direction depending on the strength of the radial flow, the
 bubble temperature, and the time when the baryons are emitted.
\end{abstract}


 \maketitle

\section{Introduction and Motivation}

Exploring the Quantum-Chromodynamics (QCD) phase diagram has been the
focus of significant experimental and theoretical research in the last
few decades. In particular, the effect of heating and compressing
nuclear matter on confinement and chiral symmetry breaking has been
studied, new phases of matter have been suggested, and the existence
of a critical endpoint.
predicted \cite{CP}.  Relativistic heavy-ion experiments also aim to
explore the temperature-baryon density plane and to provide evidence
for a critical endpoint or a first order phase transition. Therefore,
it is important to define reliable signatures for identifying a
critical point, a new phase, or a  first-order transition line.

\begin{figure}[h]
\begin{center}
\includegraphics[width=6.5cm]{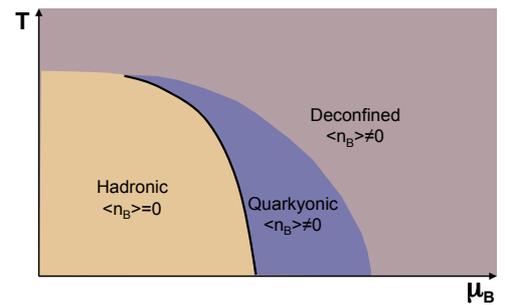}
\end{center}
\caption{Possible phase diagram of QCD.}
\label{fig:phase_diag} 
\end{figure}

In attemtping to understand QCD, it has often proven instructional to
explore QCD-like theories in which some of the parameters of the
Lagrangian (quark mass $m_q$, number of colors $N_c$, number of
flavors $N_f$) are varied \cite{thooft}. McLerran and Pisarski used
such a study to argue that at large $N_c$ and sufficiently small temperature to baryon
chemical potential ratio 
$T/\mu_B$ the phase diagram exhibits a strongly first-order phase
transition to what they termed quarkyonic matter \cite{mp}. 
For large $N_c$ and all $N_f$, baryon number has
been suggested as an order parameter for the transition \cite{hmp}: In
large $N_c$ the baryon mass is of the order $M_B \sim N_c
\Lambda_{QCD}$.  For moderate temperatures, $T \sim \Lambda_{QCD}$,
the expectation value of baryon number is $n_B \sim \exp{(\mu_B/T -
  M_B/T)} \sim e^{-N_c}$.  This is negligibly small (identically zero
at large $N_c$), and it remains that way as long as 
$\mu_B \leq M_B$.  For larger $\mu_B$ the baryon
number density becomes non-zero. In the deconfined quark-gluon plasma
phase there are no baryon masses, so that there is no baryon-number
suppression \cite{hmp}.
Figure \ref{fig:phase_diag} shows the phase diagram with a hadronic
(mesonic) phase and a high energy density quarkyonic phase. At small
temperatures the transition from the mesonic to quarkyonic phase
should be first order \cite{hmp}.  Although large $N_c$ is not QCD, it
is not excluded that the QCD phase diagram bears some resemblance to
the large $N_c$ diagram. For realistic $N_c$ and $N_f$ the boundary of
these regions may be crossovers.

The question we asked ourselves was, if there is a first-order transition between the quarkyonic and
mesonic phase, what are the observable consequences?  
The arguments laid down in the rest of the
paper are, however, more general, because they only require the
existence of a baryon-rich and a baryon-poor phase.  Phase conversion
between these can happen either by bubble nucleation or by spinodal
decomposition. 
Here we explore the phenomenological consequences of
bubble nucleation in QCD matter. Spinodal decomposition has been addressed  
in \cite{randrup}.  
It has been suggested by Voloshin \cite{voloshin} that
correlations in coordinate-space can be transferred to momentum-space
through radial flow. Baryon number fluctuations were discussed in
\cite{gavin}, but the effects of radial expansion and azimuthal
correlations were not considered in that work. We focus on the
transfer of spatial correlations between baryons confined to
baryon-rich regions, into baryon-baryon correlations in momentum space
via radial flow.

\section{Baryon-rich Bubbles}

\begin{figure}[h]
\begin{center}
\includegraphics[width=4.5cm]{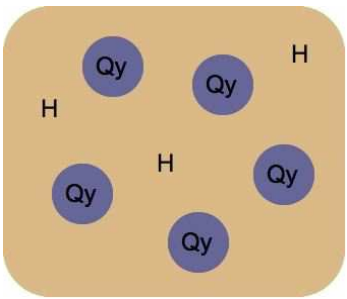}
\end{center}
\caption{Evaporating quarkyonic bubbles Qy in the hadron gas H. }
\label{fig:phases} 
\end{figure}

In this work we consider a bubble nucleation scenario, similar to the
one proposed for strangelet condensation in the early universe
\cite{witten}. Imagine an expanding system in which the initial
temperature is higher than that of the transition between quarkyonic
and mesonic matter. Initially there is non-zero baryon number density.
As the system cools a small amount below the phase transition
temperature, bubbles of stable mesonic phase start to nucleate.  
It has been shown in a dynamical calculation by Paech and Dumitru \cite{paech} that baryon inhomogeneities do form. 
The mesonic bubbles expand rapidly, reheating the system back to the
transition temperature. The system then is in a mixed phase, which at
first is mostly quarkyonic with small droplets of hadron
gas. Then these mesonic droplets increase in size.  At the time when
roughly half the matter is in the quarkyonic phase one can think of
bubbles of quarkyonic matter evaporating in the hadron gas, as shown
in Figure \ref{fig:phases}. The ratio of baryon densities of the
hadronic phase and that of the quarkyonic phase $n_H/n_{Qy} \sim
e^{-N_c}$ is tiny (identically zero in large $N_c$).  Because of this
exponential suppression of the baryon density in the mesonic phase,
all of the baryon number remains trapped inside the bubbles of
quarkyonic phase, and at later times it can concentrate in the
evaporating droplets of quarkyonic matter.

If bubbles of quarkyonic matter have baryon number larger than one,
large variation will exist in the spatial baryon number-density. In
the following we study how inhomogeneous the corresponding
distribution in momentum space will be. The longitudinal position
inhomogeneities will be translated into inhomogeneities in rapidity
because of the familiar correlation between momentum- and
coordinate-space rapidity.  In the transverse space there should be a
similar correlation due to transverse flow \cite{voloshin}.

Baryon correlations due to transverse flow are not likely to be washed
out by thermal noise. The momentum imparted to a hadron due to flow
depends on the mass and flow velocity $p_{flow} \sim M v_{flow}$ while
the thermal noise depends on the mass and temperature $p_{noise} \sim
\sqrt{MT}$. Since $p_{flow}/p_{noise} \sim v_{flow} \sqrt{M/T}$ and
since the baryon mass is at least a factor of five larger than the
temperature, we expect space inhomogeneities for baryon number to be
observable as baryon-baryon correlations in momentum space. We
investigate these by constructing a Bubble Monte-Carlo Blast-Wave
Model similar to Ref.~\cite{longacre}.

\section{Bubble Monte-Carlo Blast-Wave Model}

 Our  model has two components:
One is the Bubble Monte-Carlo, in which the spatial distribution of
baryon number and bubbles is generated; The other is a Monte-Carlo
Blast-Wave which provides the momentum distribution of baryons. We
study how spatial correlations in the source function are translated
into momentum space and 
examine the effect of baryonic bubble formation on the two-particle
correlation functions. 

In our Bubble Monte-Carlo, we first make an assumption for the average
number of baryons per bubble and the distribution of the baryon number
per bubble. We generate baryon number $N_B$ for each event according
to a gaussian distribution
\be 
dN/dN_B \sim exp\{-(N_B-\bar{N}_B)^2/2\bar{N}_B\} \, , 
\ee 
in a finite window of rapidity within which we know the average baryon
number, $\bar{N}_B$.  We then distribute the baryons into bubbles with
the number of baryons per bubble $n_B$ taking the form
\be 
dN/dn_B \sim exp\{-(n_B-\bar{n}_B)^2/2\bar{n}_B\}\, .
\ee 
We sample this baryon-per-bubble distribution to generate bubbles
until we have generated the total number of baryons within the given
window. This fixes the number of bubbles. Our treatment of $N_B$ and
$n_B$ is somewhat arbitrary, but in this analysis rather than focusing
on number fluctuations sensitive to this implementation, we will study
the shape of the two particle correlations in rapidity and azimuth
($\phi$). 
\begin{figure}[h]
\begin{center}
\includegraphics[width=9.2cm]{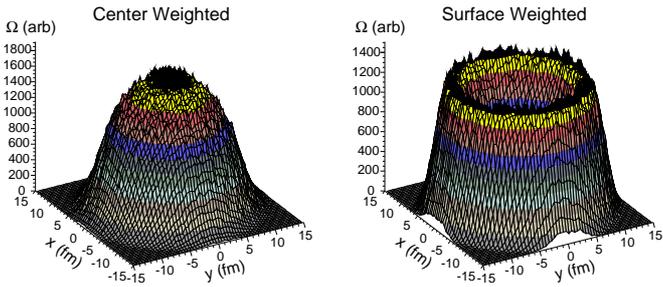}
\end{center}
\caption{The source functions with emission
  weighted most heavily from the center (left) or from the surface (right). }
\label{fig:source} 
\end{figure}

The bubbles are then distributed in space with two different transverse density profiles: The {\it Center
  Weighted Source} profile, assumes a bubble density given by a
Woods-Saxon distribution integrated along the longitudinal
z-direction. This distribution yields a density largest at the center
of the system. The {\it Surface Weighted Source} gives a preference to
bubble formation near a surface, motivated by the picture of an
expanding system where bubble nucleation happens preferentially at a
freezeout surface, near the edge of the system. Transverse projections
of both source functions are shown in Figure \ref{fig:source}. For both cases, we take the
distribution to be flat in z. Baryons are then distributed inside the
bubbles using again a Woods-Saxon distribution of the
form 
\be 
dN/dxdydz \sim 1/(1+\exp\{-(r_{b}-R)/a\})\, . 
\ee 
The bubble radius is taken to be $r_{b}=0.6~{\mbox{fm}} (n_B)^{1/3}$,
so that the volume per baryon remains constant, the radius of the
system $R=12~$fm, and the width of the distribution $a=r_{b}/8~$. In
this way we obtain the positions $(x,y,z)$ for each baryon. This discrete set of
coordinates is later taken as the source function $\Omega$.
\begin{figure}[h]
  \begin{center}
    \includegraphics[width=5cm]{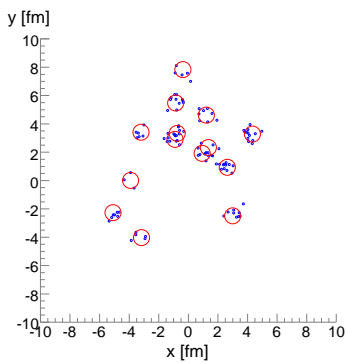}
  \end{center}
  \caption{Transverse spatial baryon- and bubble-distribution for a
    center weighted source. The circles representing the bubbles are
    drawn with a fixed radius that does not reflect the variable bubble
    size $r_{b}$. }
  \label{fig:dist} 
\end{figure}
Figure~\ref{fig:dist} shows the
baryon- and bubble-distribution in the transverse plane of one event.

Each bubble is then assigned a radial boost velocity as is done in the
blast-wave model. To construct a blast-wave Monte-Carlo we calculate
the emission function, $S$. The emission function is the probability
of emitting a baryon of transverse mass $m_T=\sqrt{p_T^2+m_B^2}$,
transverse momenta $p_T$, and rapidity $Y_B$, at an azimuthal angle
$\phi_B$ from a boosted source of $\Omega$ at temperature
$T$. 
We use the
discrete set of points generated with our bubble Monte Carlo as our
source function $\Omega$.  Following \cite{lisa}, we write the
emission function as \be
S=m_T\cosh(\eta-Y_B)\Omega(r,\phi_s)e^{-(\tau-\tau_0)^2/2\Delta\tau^2}\times \nonumber \\
\sum_{n=1}^{\infty}(-1)^{n+1}e^{na \cos(\phi_b-\phi_B)}e^{-nb \cosh(\eta-Y_B)}\, ,
\label{S}
\ee
where:  
\be
	a\equiv\frac{p_T}{T}\sinh\beta(r,\phi_s)
~~ \mbox{and} ~~
	b\equiv\frac{m_T}{T}\cosh\beta(r,\phi_s)\, .
\ee
Here we will consider central collisions, which implies the direction of the
boost $\beta$ is identical to the spatial azimuthal angle $\phi_b=\phi_s$.
Generalization to finite impact parameter presents no major technical
difficulties. We allow the bubbles to
decay isotropically in their flow boosted rest frame.  Assumption of boost invariance manifests in the equality of
the longitudinal flow rapidity to the space-time rapidity $\eta$. 
For simplicity, we assume that the emission is
instantaneous and replace the Gaussian in (\ref{S}) with 
$\delta(\tau-\tau_0)$. We
are left to specify the transverse rapidity boost $\beta=r[\beta_0+\beta_2\cos(2\phi_b)]$. The parameter $\beta_0$ is the transverse rapidity in the
outward $\phi_s$ direction. We take
$\beta_2$ to be zero. We will
investigate different values for $\beta_0$ taking guidance from the fits 
in \cite{lisa}, which reproduces well the experimental data.

\section{Baryon-Baryon Correlation Results}
\label{results}

We investigate how baryon-baryon correlations in
terms of relative angle (azimuth $\Delta\phi$ and rapidity $\Delta y$)
vary with the freeze-out temperature $T$, average flow velocity
$\langle\beta\rangle$, number of baryons per event $N_{B}$, number of
baryons per bubble $n_B$, time at which bubbles decay $\tau_{decay}$, and source
function (Surface vs Center Weighted). We present the correlations in
terms of a per particle correlation measure \cite{trainor}
\be
\frac{\Delta\rho}{\sqrt{\rho_{ref}}} = \frac{d^2N/(d\phi_1d\phi_2) - (dN/d\phi_1)\cdot(dN/d\phi_2)}{\sqrt{ (dN/d\phi_1)\cdot(dN/d\phi_2)}}\, .
\ee
Here $\rho$ is the pair density and $\rho_{ref}$ the product of single
particle densities. This normalization of the correlation measure
means that for all other parameters kept fixed
$\Delta\rho/\sqrt{\rho_{ref}}$ is independent of the total number of
baryons per event $N_B$ \cite{trainor}. This is because normalizing by
$\sqrt{\rho_{ref}}$ instead of $\rho_{ref}$ removes the dilution
factor of one over the number of particles analyzed. This was
confirmed to be true in our Monte-Carlo (the variation of $r_b$ with $N_B$ does not seem to influence the correlation function). We generate our reference
spectrum by sampling the single particle $dN/d\phi$ and $dN/dy$
distributions generated in our Monte-Carlo enough times so that the
same number of pairs are created in our reference distribution as in
our real distribution. This ensures that
$\Delta\rho/\sqrt{\rho_{ref}}$ will integrate to zero. For this reason
we are insensitive to a global offset and focus instead on the shape
of the correlation, which is easier to measure experimentally.
\begin{figure*}[htbp]
  \begin{center}
    \includegraphics[width=5cm]{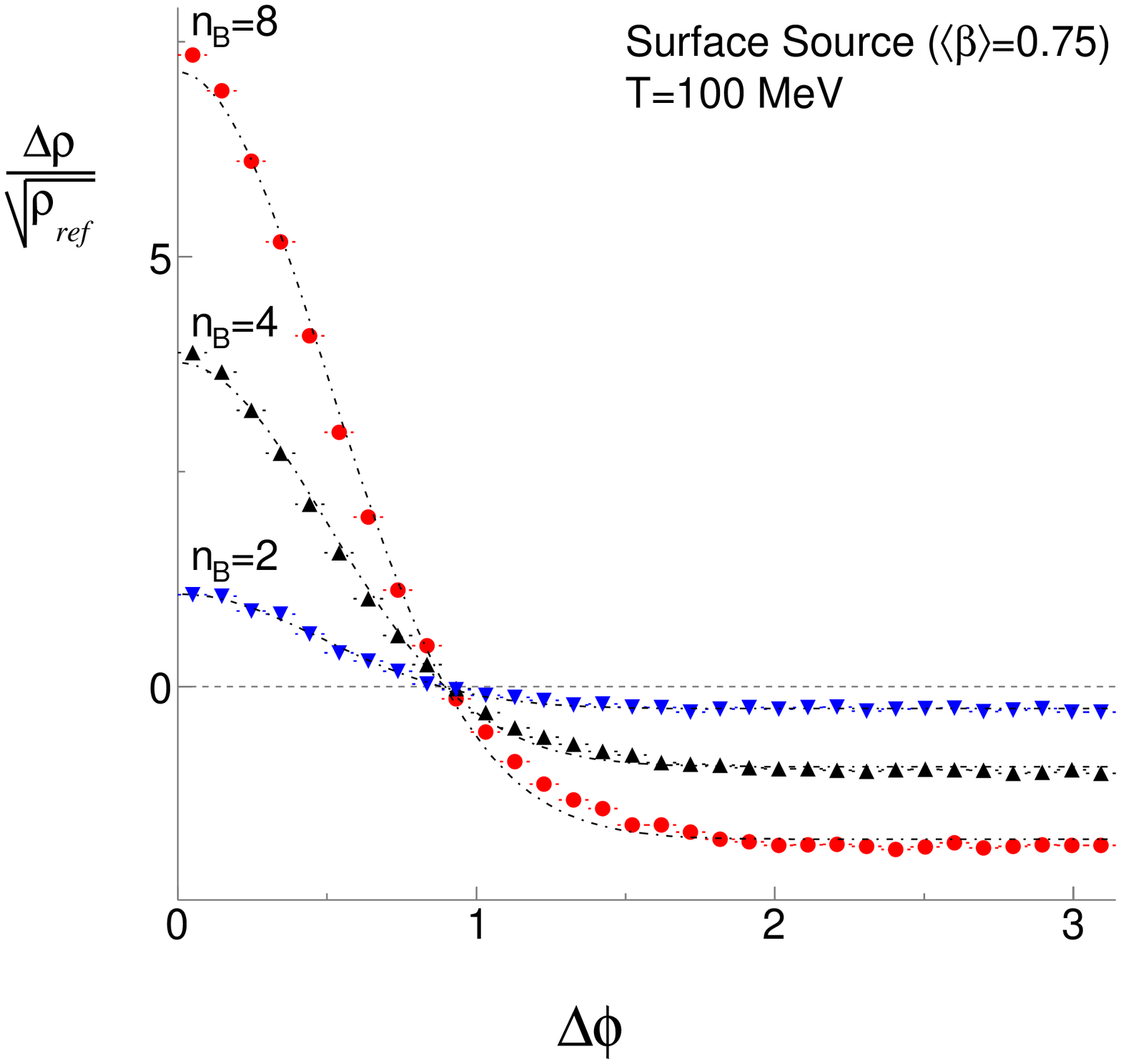}
    \includegraphics[width=5cm]{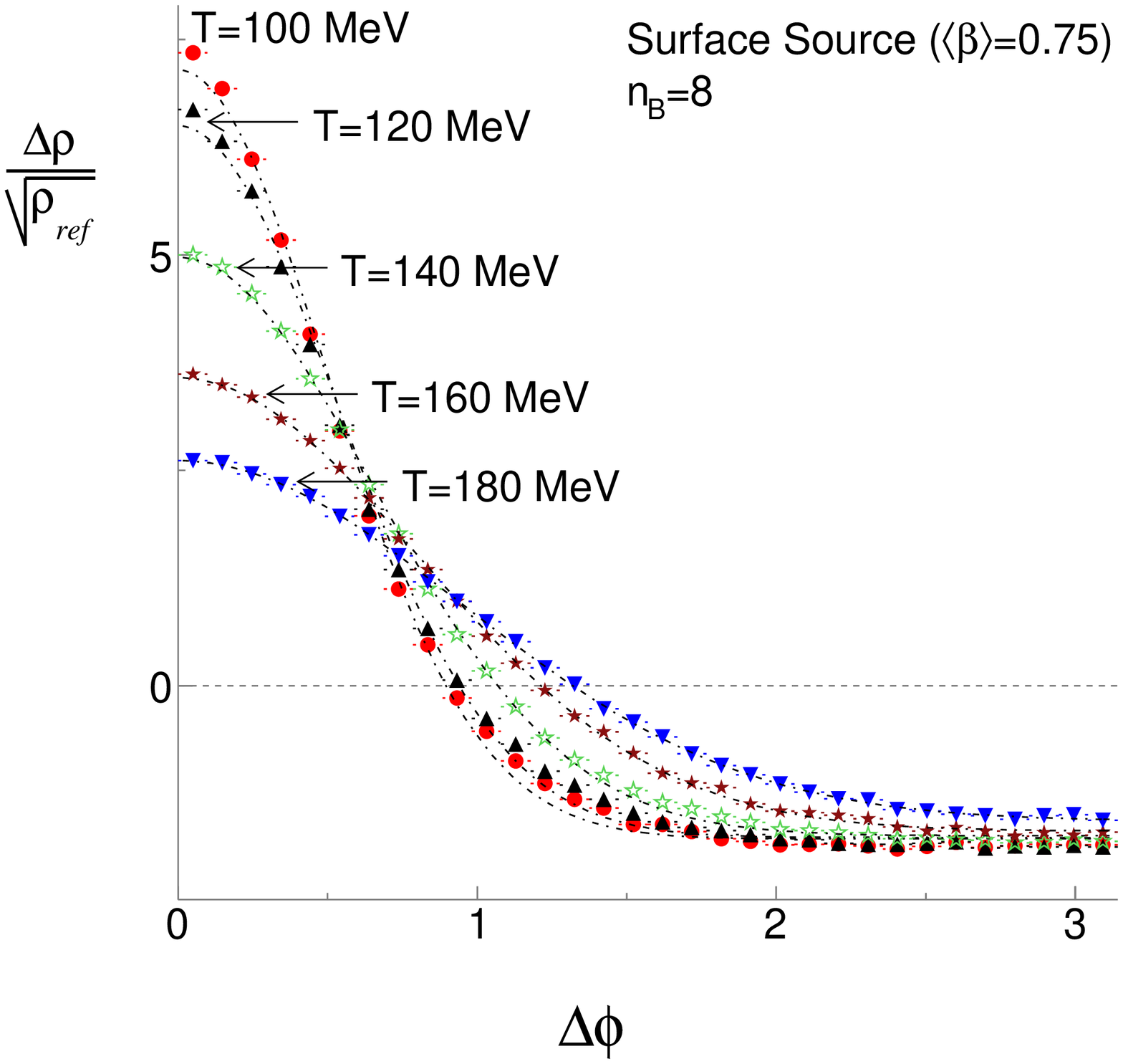}\\
    \includegraphics[width=5cm]{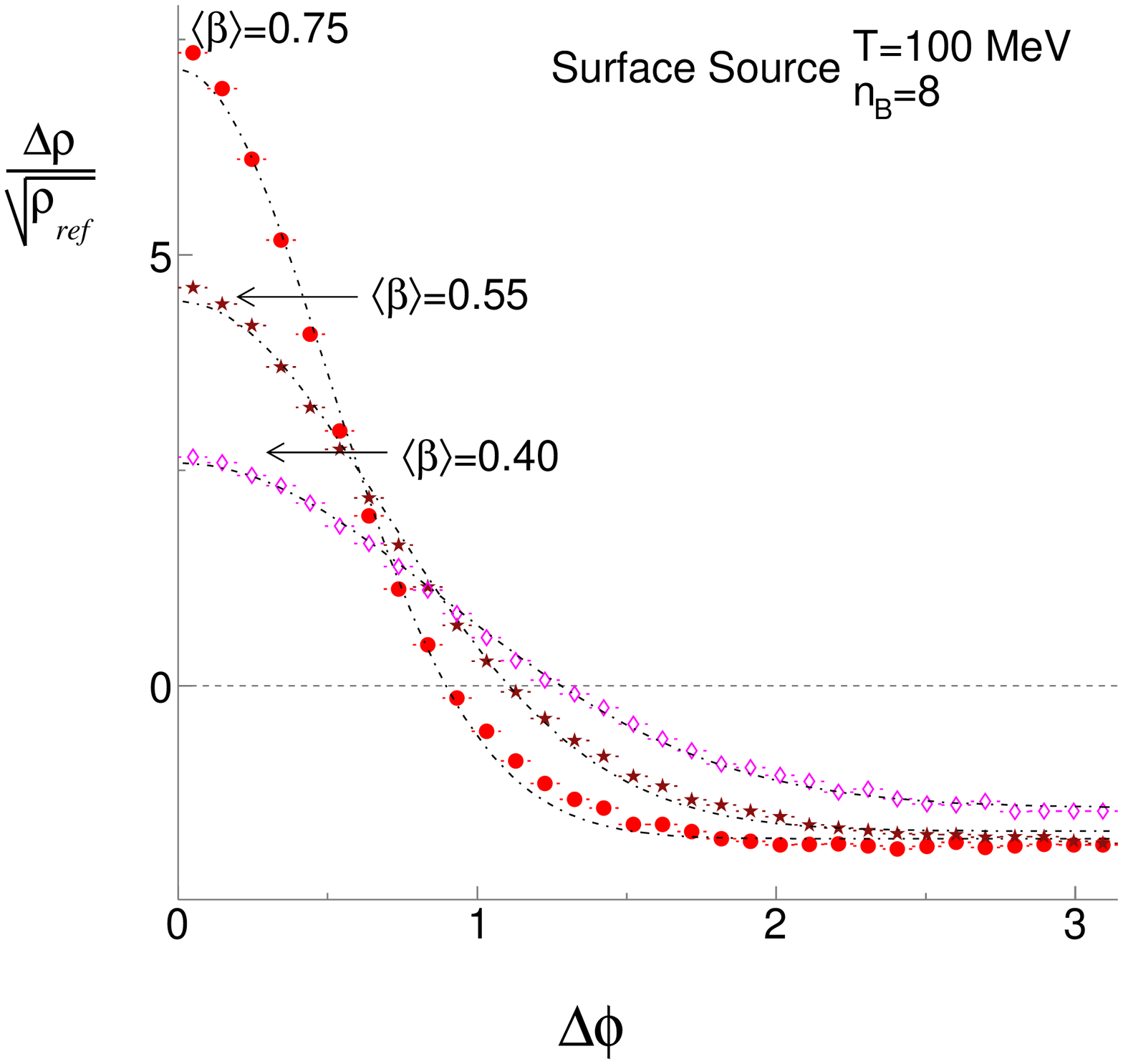}
    \includegraphics[width=5cm]{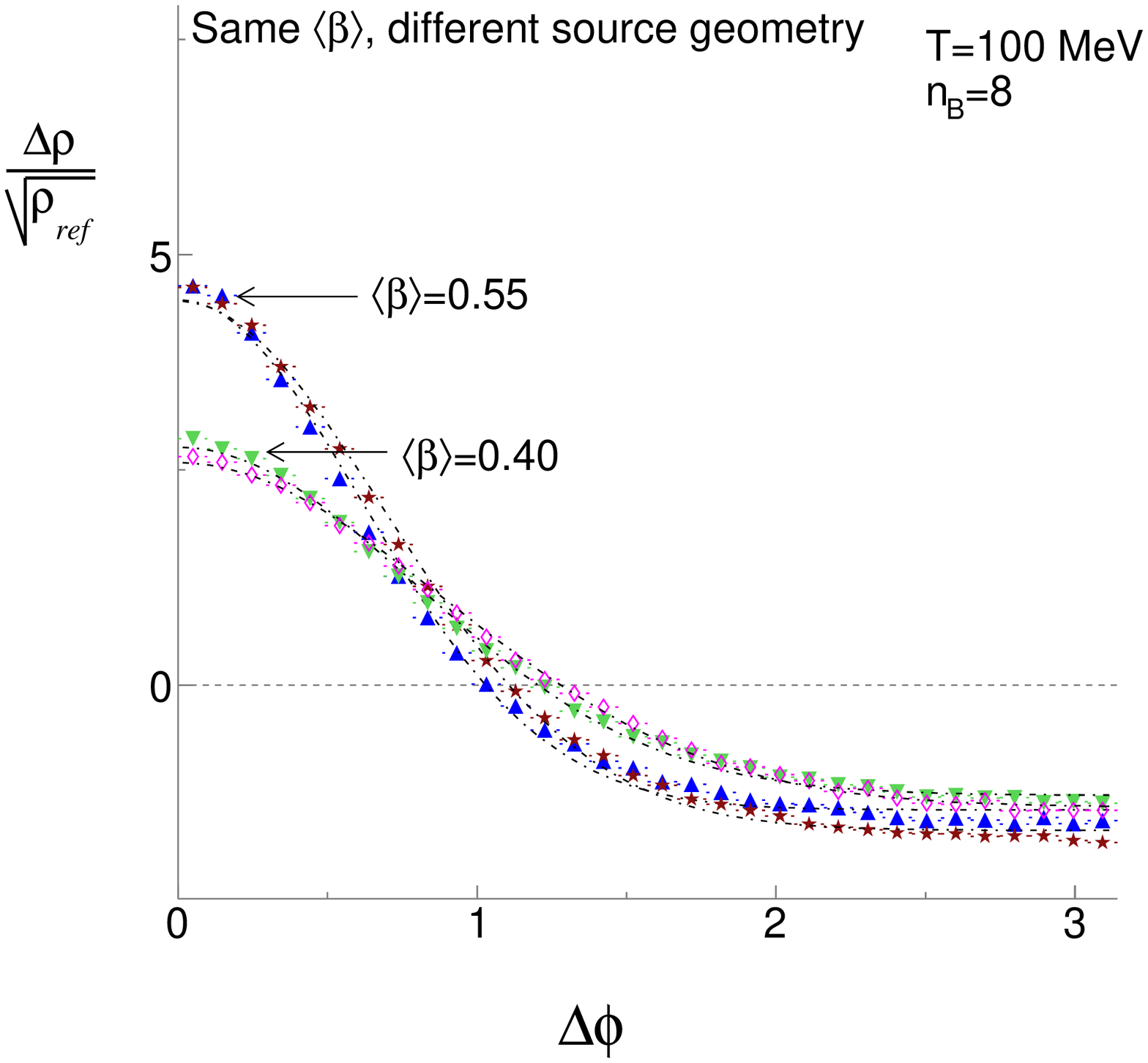}
  \end{center}
  \caption{Proton-proton correlations in terms of azimuthal angle
    between them for different model parameters. For details see the
    text.  }
  \label{fig:result} 
\end{figure*}
The variable $\Delta\rho/\sqrt{\rho_{ref}}$ as a function of the
azimuthal angle between baryons from the decayed bubble for a variety
of different model inputs is shown in Figure~\ref{fig:result}. Bubble
nucleation and radial flow leads to small angle correlations.  We find
that all correlation shapes can be reasonably well parameterized by a
Gaussian centered at $\Delta\phi=0$ and a constant offset. The
correlation can therefore be summarized in terms of the Gaussian
amplitude $A$ and root mean squared width $\sigma$. By construction
the quantity $\Delta\rho/\sqrt{\rho_{ref}}$ is independent of the
total number of baryons per event \cite{trainor}. We confirmed this in
our simulation, and then fixed $N_B=100$.

The top left panel in Figure~\ref{fig:result} shows results with a
Surface Source and $T=100~$MeV, 
$\langle\beta\rangle=0.75$, and $n_B=2, 4, 8$. The width of the correlation is narrowest
for $n_B=2$, since this corresponds to the smallest bubble size, but
the variation of $\sigma$ with $n_B$ is very weak (only changing from
0.487 for $n_B=2$ to 0.499 for $n_B=8$). The dominant effect of
varying $n_B$ is on the amplitude of the correlation, which changes
from 1.3 to 8.9, respectively. This dependence can be understood in
the following way: The number of pairs that can be formed from one
bubble is $n_{B}(n_B - 1)/2$. So the number of correlated pairs in an
event is $n_{B}(n_B - 1)/2$ times the number of bubbles, which is 
the number of baryons $N_B$
divided by the number of baryons per bubble $n_B$. Thus the number of
correlated pairs in an event is $N_B(n_B - 1)/2$. Recall, that the
variable $\Delta\rho/\sqrt{\rho_{ref}}$ measures the number of
correlated pairs per particle. For our simulation
$\frac{\Delta\rho}{\sqrt{\rho_{ref}}} \propto \frac{N_B(n_B -
  1)}{\sqrt{N_B^2}} = n_B-1$.  Thus the correlation
amplitude scales with $n_B-1$. Accordingly, the $n_B = 8$ amplitude is
approximately 7 times larger than for $n_B = 2$. We note here that
this amplitude only applies to the case that the correlation is formed
only with baryons produced from bubbles. Any background contribution
(for instance from baryons produced outside of bubbles) will dilute the signal.

In the top right panel in Figure~\ref{fig:result} we show the
dependence of the correlation on the temperature of the bubbles
emitting the baryons, for a fixed $\langle\beta\rangle=0.75$ and
$n_B=8$. This temperature is not necessarily the temperature at which
the bulk of the hadrons freeze-out, but could be
closer to the value of the critical temperature $T_c$, where the
bubbles actually form. We vary this T from $120~$MeV to
$180~$MeV. It is clear that the shape of the correlation changes
appreciably with temperature: higher T lead to a broader
correlation function.  In this temperature range and for this flow
velocity the r.m.s. width is given by the fit $\sigma =
0.694 - 0.00639T + 0.0000435T^2$.  Since the
random thermal boost competes with the flow boost, a larger bubble decay
temperature washes out the correlation. Therefore, if the first order
phase transition is at a high $T_c$, correlations from bubbles will be
harder to detect.

Since we expect a competing effect of the radial flow and the
temperature, we now look at how flow influences the correlation. We
illustrate this for  $T=100~$MeV, $n_B=8$, and $\langle\beta\rangle=0.75, 0.55, 0.40$ on the bottom
left panel of Figure~\ref{fig:result}. The correlation is
strongly dependent of the mean flow velocity and is the 
strongest for bubbles with larger boost. Remember, that flow is
required to convert coordinate-space correlations to momentum
space. So again, if the phase transition happens early, before flow
builds up, bubbles will be hard to
detect. 

The bottom right panel of Figure~\ref{fig:result} displays the flow
effects at $T=100$ and $n_B=8$, with either a Center- or
Surface Weighted source function. The
mean radial position of the source is larger for the Surface Weighted
source: $10.1~$fm compared to $7.4~$fm. This leads to a larger average
boost velocity: $\langle\beta\rangle=0.75$ compared to
$\langle\beta\rangle=0.55~$, respectively. A larger
$\langle\beta\rangle$ leads to a more pronounced correlation structure
with the Gaussian narrowing from $\sigma = 0.61$ to $\sigma =
0.50$. We also compared a Surface Weighted source with a reduced
$\beta_0$ such that the $\langle\beta\rangle$ is matched to the Center
Weighted case. The Surface Weighted correlation with
$\langle\beta\rangle=0.55$ is only slightly narrower than the Center
Weighted correlation with the same $\langle\beta\rangle$. The primary
difference between Surface- and Center Weighting is caused by the
change in $\langle\beta\rangle$, but some deviation occurs due to the
difference in geometry. Our interpretation is that the Surface
Weighted source has a smaller range of $\beta$ values, while the
Center Weighted source has a broad distribution of $\beta$ values. In
this case, even for the same $\langle\beta\rangle$ value, the Center
Weighted source gives a convolution of a broad range of $\beta$ values
which leads to a broader correlation function. Also, a bubble near the
center covers a larger solid angle. We investigated
three $\langle\beta\rangle$ values for each geometry. The widths
within the ranges studied can be parameterized as follows: $\sigma =
1.317 - 1.103\beta$ for the Center Weighted source (see lower right
panel) and $\sigma = 1.357 - 1.367\beta$ for the Surface Weighted
source (see lower left panel). These parameterizations of our model
are valid for $0.29<\beta<0.75$.

\begin{figure}[htbp]
  \begin{center}
    \hspace*{-.6cm}
    \includegraphics[width=5cm]{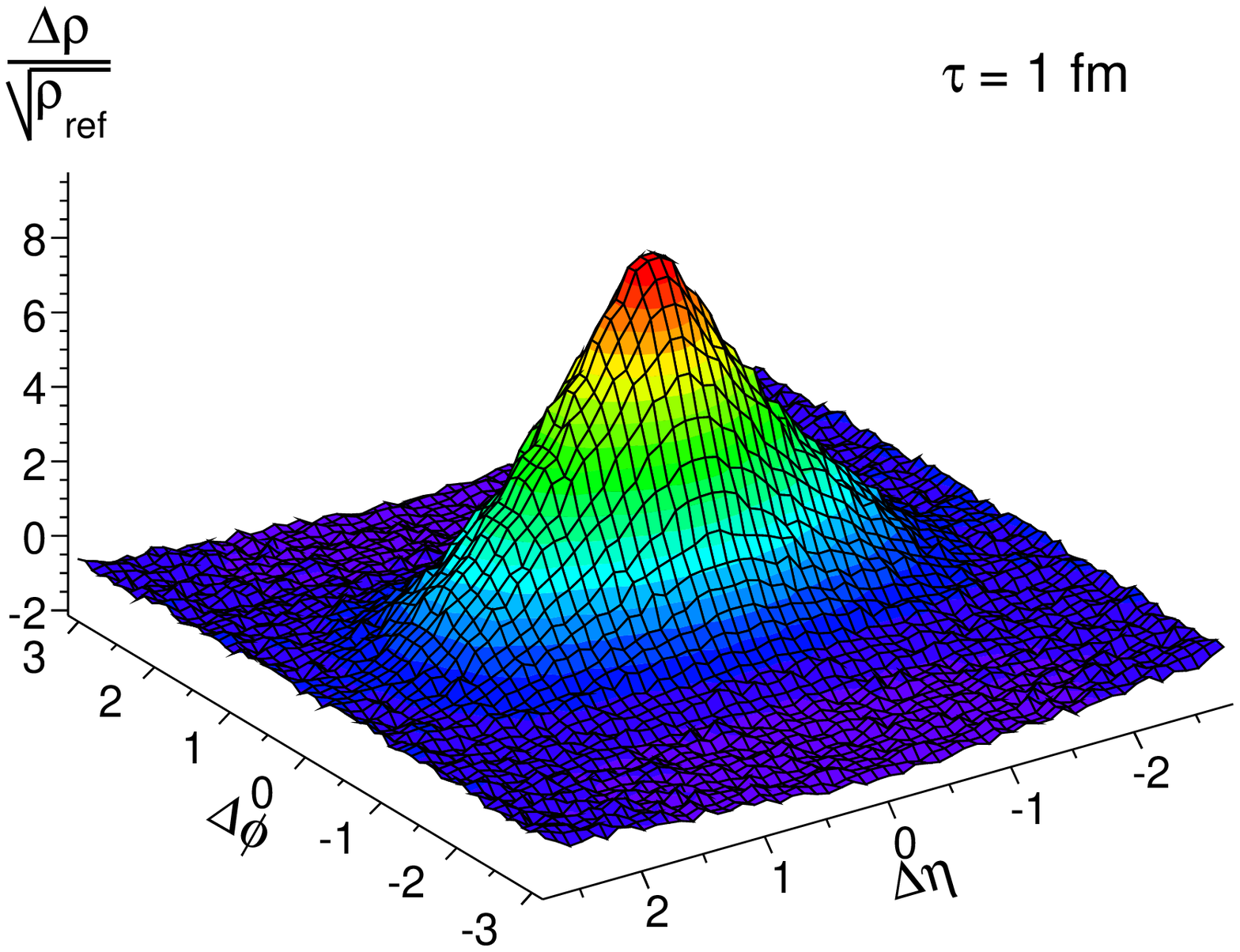}
    \hspace*{-0.9cm}
    \includegraphics[width=5cm]{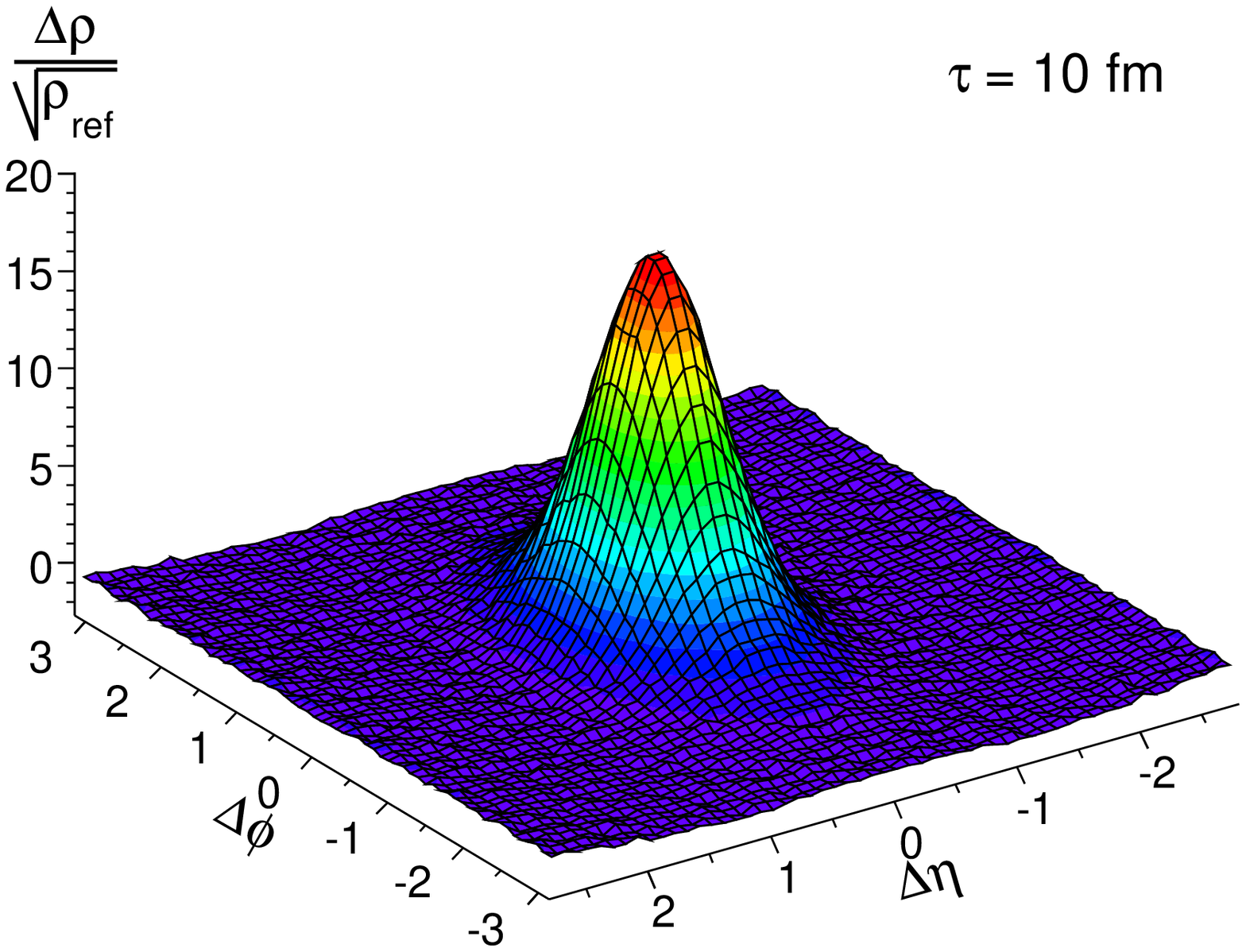}
  \end{center}
  \caption{Proton-proton correlation versus azimuthal angle and
    rapidity difference between the protons for bubble decay at times
    times $\tau_{decay}=1~$fm (left) and $\tau_{decay}=9~$fm
    (right). }
  \label{fig:rap} 
\end{figure}
We also looked at correlations between baryons in the rapidity $y$
direction. Figure \ref{fig:rap} shows the proton-proton correlation as a
function of rapidity difference and azimuthal angle between the
protons at different proper times of bubble decay, i.e. $\tau_{decay}=1~$fm
(left panel) and $\tau_{decay}=9~$fm (right panel). We found that if
the time at which bubbles decay $\tau_{decay}$ is long then narrow correlations can build up
both in $\Delta\phi$ and $\Delta y$ directions. These correlations are
increased in amplitude and narrowed in width. We conclude that if
bubble decay happens near the beginning of the system evolution (small $\tau_{decay}$) 
implying a high decay temperature and low flow, baryon-baryon
correlations will be broad in both directions. For bubble decay near
freeze-out (large $\tau_{decay}$), on the other hand, strong correlations with narrow widths
in both directions can be expected. Our fit for the change
in the width in rapidity direction with time is
$\sigma_y=0.37+0.47\tau^{-1.58}$.

Our studies of $\Delta\rho/\sqrt{\rho_{ref}}$ show that, the amplitude
and width of the correlation varies substantially within this bubble
Monte-Carlo blast-wave model depending on the parameters assumed: We
have shown in Figure~\ref{fig:result} That correlation caused by
baryonic bubbles emitted from a boosted source can change by a factor
of two even with reasonable values for the average boost velocity and
freeze-out temperature.  The signature of a first order phase
transition will be strong if the transition happens near the freeze
out since then the flow will be largest and the temperature lowest. We
note, that the decay time is linked to the temperature and flow, and
with a dynamical model one could make this connection explicit.

We also consider how this correlation can affect measurements of
elliptic flow $v_2$~\cite{flow} and it's fluctuations \cite{psorensen}. One
method for estimating $v_2=\langle \cos(2\phi-\Psi)\rangle$ when the
reaction plane angle $\Psi$ is not known is to calculate
$\langle\cos(2\Delta\phi)\rangle$ from two-particle correlations. This
quantity depends on $\langle v_2\rangle^2$, $\sigma_{v_2}^2$, and
other correlations not related to the reaction plane, called non-flow
$\delta_2$. Since we only consider central collisions in this study
$\langle v_2\rangle$ is zero. The calculation of $\langle
cos(2\Delta\phi)\rangle$ is therefore related to either non-flow
correlations or $v_2$ fluctuations. In either case, the correlations
from our model will lead to a deviation between the $2^{nd}$ and
$4^{th}$ order cumulants~\cite{flow} such that $v_2\{2\}^2-v_2\{4\}^2$
is non-zero. Since $v_2$ is zero the expected deviation between the
cumulants is given by
\be v_2\{2\}^2-v_2\{4\}^2 =
\frac{\int_0^\pi d\Delta\phi \cos(2\Delta\phi)dn_{pairs}/d\Delta\phi}{\int_0^\pi d\Delta\phi dn_{pairs}/d\Delta\phi }\, .
\ee
This quantity depends strongly on the variables explored in the
discussion above, as well as on the number of uncorrelated particles
included in the analysis (misidentifed mesons for example). Larger
multiplicities will dilute the correlation leading to smaller
differences between $v_2\{2\}^2$ and $v_2\{4\}^2$. The quantity
$N_{B}(v_2\{2\}^2-v_2\{4\}^2)$, however, should be independent of
$N_B$ if all other paremeters are fixed because. The factor of $N_{B}$
cancels out the dilution of $v_2\{2\}^2-v_2\{4\}^2$ arising from the
increase of the number of possible pairs when the multiplicity
increases. To confirm this we kept other parameters fixed and varied
$N_{B}$ from 60 to 200. The resulting $N_{B}*(v_2\{2\}^2-v_2\{4\}^2)$
did not change. With $T=100$~MeV and $\langle\beta\rangle = 0.55$, we
found that $N_{B}(v_2\{2\}^2-v_2\{4\}^2) \approx 5$ for $n_B=8$ and $
\approx 2.7$ for $n_B=4$. Increasing the temperature to $180~$MeV with
$n_B=4$ reduces $N_{B}(v_2\{2\}^2-v_2\{4\}^2)$ to $ \approx 0.37$.
The introduction of a number $M$ of uncorrelated background particles
(pions for example), reduces $N_{B}(v_2\{2\}^2-v_2\{4\}^2)$ by a
factor of $M^2$. Putting everything together we find
$v_2\{2\}^2-v_2\{4\}^2 =\frac{C}{N_BM^2}$,
where $C$ can be expected to be between 0.3 and 5.

Measurements of p-p and p-$\Lambda$ correlations have
been carried out at energies ranging from $\sqrt{s}\sim2$ GeV
to 200 GeV \cite{Barrette:1999qn}. In
these HBT analyses correlation functions are defined as
$C(k^*)=N_{pair,real}/N_{pair,mixed}$.  The variable 
$k^*=Q_{inv}/2$ is the relative momentum
of the particles in the pair rest frame, $N_{pair,real}$ is the number
of pairs in an event with relative momentum $k^*$, and $N_{pair,mixed}$
is the number of pairs expected from random combinatorics and is
constructed from mixed events. The region of interest to an HBT
analysis is below $k^* \sim 100~$MeV. The $N_{pair,mixed}$ distribution
is usually normalized to the $N_{pair,real}$ distribution at some
value of $k^*$ above this region.  Our Monte Carlo model yields a
correlation $C(k^*)$ well described by a Gaussian with an RMS width of
750 MeV. For bubbles with $n_B=8$ the
amplitude of the Gaussian is 8\%. If the $N_{pair,mixed}$ distribution
is normalized to the region near $k^*=500~$MeV the correlation induced
by bubble formation would show up as a smooth 1--2\% variation of
$C(k^*)$ between 0 and $500~$MeV. In HBT analyses, long-range
non-femptoscopic correlations of this kind are treated as background
and ignored or parameterized. We therefore conclude that although HBT
measurements performed at lower beam energies do not report signs of a
strong baryon-baryon correlation as described here, they also do not
exclude their presence.

\section{Summary and conclusion}

We have shown that bubble nucleation at a first or phase transition
coupled with radial flow in heavy-ion collisions can lead to
detectable baryon-baryon correlations. We've mapped out the shape of
these correlations in rapidity and azimuthal angle and discussed how
they will manifest as non-flow or flow fluctuations by contributing to
the difference between $v_2\{2\}$ and $v_2\{4\}$. We explored how the
correlations will depend on the strength of the radial flow, and the
temperature of the system at the time when the baryon rich regions
decay. We reported the variation of the azimuthal and longitudinal
width with temperature, flow, geometry, and decay time. We find that
if bubble decay happens late in the evolution, close to the freeze-out
when the temperature is lower and the flow is larger, the correlations
will be narrow in rapidity and azimuth. We conclude that these
correlations would be easy to detect. If the bubbles decay early in
the evolution, the flow will be weaker, the temperature will be
higher, and the correlation will be wider in both directions and
therefore less pronounced in the data. In this later scenario, there
will be more hadronic rescattering which may wash out the signal
entirely. The effect of the hadronic stage on these correlations is an
important study which will be the focus of future more quantitative work.

The observation of baryon-baryon correlations as described in this
work will be evidence for the existence of a first-order phase
transtion. The lack of such a signal will indicate that either bubble
nucleation at a first order phase transition did not occur, or that
the transition was sufficiently seperated from freeze-out as to render
the correlations unobservable. We propose therefore, that studies of
baryon-baryon correlations can be used to answer whether a first-order
phase tranition is present in the vicinity of the freeze-out curve in
heavy-ion collisions. Future studies of the effect of the hadronic
rescattering stage will allow us to more precisely specify the region
within which a first order phase transition should be detectable. If
the transtion to quarkyonic matter is first order and close enough to
hadronic freeze-out, then we conclude that it can be
detected through baryon-baryon correlations.

\section*{Acknowledgments}

W thank L.~McLerran for his collaboration at the early stages of this
work, and R.~Pisarski, E.~Fraga, G.~Krein, L.~Palhares, T.~Kodama for
discussions and support.


\end{document}